\documentclass[preprint]{aastex}

\def\newpage{\vfill\eject}   
\newcommand{\be}{\begin{equation}}
\newcommand{\ee}{\end{equation}}
\newcommand{\mpl}{ {M_{\rm pl}}} 
 
\newcommand{\lpl}{ {\ell_{{\rm pl}}}} 
\newcommand{\dist}{ { d_\ast }} 
\newcommand{\egam}{ { E_\gamma }} 
\newcommand\pp{\parshape 2 0.0truecm 16.0truecm 1.5truecm 14.5truecm}

\begin{document}

\baselineskip=20pt 
\parskip=1.0ex 

\title{\bf POSSIBLE LIMITS ON PHOTON PROPAGATION \\ 
FROM QUANTUM GRAVITY AND SPACE-TIME FOAM} 

\author{Manasse Mbonye$^1$ and Fred C. Adams$^2$} 

\affil{$^1$Physics Department, Rochester Institute of Technology, 
Rochester, NY 14623}  

\affil{$^2$Michigan Center for Theoretical Physics, Univ. Michigan, 
Ann Arbor, MI 48109} 
 
\begin{abstract} 

Many quantum gravity theories imply that the vacuum is filled with
virtual black holes. This paper explores the process in which high
energy photons interact with virtual black holes and decay into
gravitons and photons of lower energy. The effect requires violation
(or modification) of Lorentz invariance and implies that high energy
photons cannot propagate over arbitrarily large distances. For the
standard Planck mass and the likely form for the interaction cross
section, this quantum foam limit becomes $\dist < 450$ Mpc
$(\egam/10^7 {\rm GeV})^{-5}$. For quantum gravity theories that posit
a lower Planck scale, the interaction rate is larger and the limit is
stronger. This paper uses extant observations of gamma rays from
cosmological sources to constrain this process for varying values of
the Planck mass and a range of forms for the interaction cross
sections.

\end{abstract} 

constraints 
\bigskip  

Quantum gravity currently lacks definitive experimental tests. 
However, most theories of quantum gravity predict that space is filled
with virtual black holes, which can absorb photons and re-radiate them
as the black holes evaporate. If energy and momentum are conserved
during such an interaction, the virtual black hole will usually
re-radiate a photon with the same energy and direction as the original
(absorbed) photon. In general, the phase of the emitted photon will be
different from that of the absorbed photon and the photon is delayed
by a small time interval $\Delta t \sim \mpl^{-1}$, but these
processes do not affect most observations. In addition to phase
changes and time delays, however, virtual black holes can also radiate
multiple particles, provided that the relativistic dispersion relation
has a modified form (as predicted by many versions of quantum
gravity). In spite of its lower probability, this latter effect is
more readily observable and can be used to constrain theories of
quantum gravity. This paper uses existing observations of high energy
photons to place constraints on this process.

\newpage 

For astronomical sources that are close enough so that cosmic
expansion can be neglected, the optical depth $\tau$ for photons to
interact with virtual black holes takes the simple form 
\be
\tau = n \sigma \dist  \, , 
\ee
where $n$ is the number density of virtual black holes, $\sigma$ is
the interaction cross section, and $\dist$ is the distance from the
astronomical source. (This expression is generalized below to include
cosmic expansion). A successful observation of an astronomical source
implies that $\tau \le 1$.

According to the scenario of virtual black holes filling the vacuum --
the space-time foam -- the vacuum contains about one Planck mass black
hole per Planck volume [1--3]. The number density of virtual black
holes thus takes the form 
\be
n = \alpha \mpl^3 \, , 
\ee
where $\mpl$ is the Planck mass and $\alpha$ is a dimensionless
constant of order unity. [We use units in which $\hbar$ = 1, $c$ = 1,
$G = \mpl^{-2}$, and the Planck length $\lpl = \mpl^{-1}$.] 

Next we need to specify the cross sections for photons interacting
with virtual black holes. The geometrical cross section is $\sigma_0
\approx \pi \lpl^2$. Since all photons of astronomical interest are in
the long wavelength regime $\lambda \gg \lpl = \mpl^{-1}$, the
absorption cross section $\sigma_1$ is highly suppressed relative to
$\sigma_0$.  Large black holes are thought to emit radiation with a
nearly thermal spectrum [4]. If perturbed virtual black holes act 
similarly and the absorption cross section is the same as the emission
cross section, the long wavelength limit of the absorption cross section 
takes the form $\sigma_1 = \beta_1 \pi \lpl^2 (\lpl/\lambda )^2$ =
$\beta_1 \pi \egam^2 \mpl^{-4}$, where $\beta_1$ is a dimensionless
parameter and the energy $\egam=\lambda^{-1}$. For classical photon 
fields interacting with a static, uncharged, non-rotating
(Schwarzschild) black hole, the absorption cross section has been
calculated [5]; in the long wavelength limit, one finds $\beta_1$ =
64/3 (although quantum effects could modify this value).  This
absorption cross section implies an optical depth $\tau_1 = \alpha
\beta_1 \pi d_1 \lpl \lambda^{-2}$.  The path length $d_1$ required
for $\tau_1 > 1$ takes the form $d_1 \approx$ 1 Mpc $(\lambda / 1
\mu{\rm m})^2$ and is thus astronomically interesting for optical
photons. In most cases, however, the virtual black hole will emit a
photon with the same energy (but with a different phase) and the
absorption event would be impossible to detect.

Here we consider the case where the absorption of a photon by a
virtual black hole leads to the emission of two particles rather than
one. In order to conserve energy and momentum, the outgoing particles
must travel parallel to the incoming photon.  When a photon (spin-1)
emerges, the second particle must be a graviton (spin-2) to conserve
spin angular momentum. In conventional particle physics, this process,
sometimes called photon splitting, is not generally allowed for two
reasons: (a) The phase space for the outgoing particles vanishes
because the momenta are all parallel, and (b) The amplitude vanishes
because the contractions of the momenta with each other (or with the
polarizations) vanish [6]. However, the process is allowed if it
violates (or modifies) Lorentz symmetry. For example, the dispersion
relation for massless particles could have an additional term [7],
$\egam^2 (p) = p^2 + \xi p^{n+2}/\mpl^n$. In order for the phase
space to have non-vanishing volume, at least one of the outgoing
particles must have a Lorentz-violating (or modifying [7]) factor
$f\sim\xi(p/\mpl)^n$; this same factor allows the matrix elements to
be nonvanishing as well.  As a result, we expect the cross section
$\sigma_2$ for photon absorption and re-radiation of two particles to
take the general form 
\be 
\sigma_2 = \beta_2 \pi \lpl^2 (\lpl / \lambda)^b 
= \beta_2 \pi \egam^b \mpl^{-(b+2)} \, , 
\label{eq:crossb} 
\ee 
where $\beta_2$ is a dimensionless constant. Given the present
uncertainties, the index $b$ is left as a free parameter; however, 
a simple phase-space argument suggests a lowest order value of 
$b$ = 5. 

With the number density and cross section specified, the optical 
depth for two particle down-scattering takes the form  
\be
\tau_2 = \alpha \beta_2 \pi (\dist/ \lpl) (\egam/\mpl)^b \, . 
\ee 
This result can be expressed in terms of the path length required for
the optical depth to exceed unity. As photons travel across the
universe, they will experience a quantum foam cutoff at a distance
scale $\dist$ = $(\lpl/\pi) (\mpl/\egam)^b$. For example, if we take
$\alpha = 1 = \beta_2$, $b=5$, and $\mpl \approx 10^{19}$ GeV, this
quantum foam cutoff becomes 
\be 
\dist \le \, 450 \, {\rm Mpc} \, (\egam / 10^7 {\rm GeV})^{-5} \, .  
\label{eq:cutoff}
\ee 

For path lengths that are comparable to the cosmological horizon
scale, one must take into account the expansion of the universe and the
redshifting of photons with cosmological time. With this
generalization, the optical depth takes the form 
\be
\tau_2 = {\alpha \beta_2 \pi c \over \lpl H_0} \, 
\Bigl( {\egam \over \mpl} \Bigr)^b \int_a^1 
{a^{1/2} da \over a^b \, \bigl[ \Omega_M + \Omega_V a^3 
\bigr]^{1/2} } \, , 
\label{eq:cosint} 
\ee
where $H_0$ is the Hubble constant and $a$ is the cosmic scale factor.
We have assumed a spatially flat universe with matter density
$\Omega_M$ = 0.3 and constant vacuum energy density $\Omega_V$ = 0.7,
in concordance with current observations [8].

The Planck mass can be lower than its standard value $\mpl \approx
10^{19}$ GeV. Many recent papers [9] explore the possibility of a
smaller scale for quantum gravity (lower Planck mass) and larger extra
dimensions in string theory. While these theories have 10 or 11
space-time dimensions, the calculation of space-time foam continues to
predict one virtual black hole per Planck volume [10].  The quantum
foam cutoff constructed in this letter can be used to constrain the
value of the Planck mass in this context. The cross section depends
sensitively on the Planck scale (eq. [\ref{eq:crossb}]) so that
interactions of photons with virtual black holes become far more
likely with a lower Planck mass.

Figure 1 shows the maximum propagation distance as a function of
photon energy $\egam$ using the interaction cross section with $b$=5
and varying values of the quantum gravity scale $\mpl$. Because high
energy photons have already been observed from astronomical sources
[11,12], a portion of the plane is already known to be unaffected by
quantum foam; this region is shown as the shaded part of the plane
(see also Refs. [13,6]). In addition to possible interactions with
virtual black holes, high energy gamma rays can interact with photons
from the radiation backgrounds of the universe. Gamma rays with
energies $E > 300$ TeV can scatter off photons from the cosmic
microwave background radiation (CMB) and produce $e^+ e^-$ pairs. The
mean free path for pair production is only 10 kpc for photons above
the energy threshold [14]; this bound is shown as the dashed
horizontal line in Figure 1.  For the standard value of the Planck
mass ($\mpl \approx 10^{19}$ GeV), the bound from quantum foam becomes
more restrictive (with a path length less than 10 kpc) than that due
to the CMB for photon energies $\egam > 10^8$ GeV. Another bound
arises from interactions with the cosmic background of infrared
photons. The number density of photons in the infrared background is
smaller than that of the CMB, and the energy threshold is lower (30
TeV).  For photon energies $\egam > 30$ TeV, the mean free path is
about 1 Mpc [14]; this bound is shown as the dotted horizontal line in
Figure 1.  For the standard value of the Planck mass, the bound from
quantum foam is more restrictive than that from the infrared
background for $\egam > 3 \times 10^7$ GeV.

These quantum foam bounds become stronger for lower values of the
Planck mass $\mpl$. Figure 1 shows that existing data rule out quantum
gravity scales lower than $\mpl \sim 10^{15}$ GeV for cross sections
with $b=5$. A more general bound can be obtained from the entire
$b-\mpl$ plane. Observed high energy photons from extragalactic
sources (with given energy $\egam$ and known distance $\dist$) must
have $\tau_2 < 1$ and imply a limit on the Planck mass as a function
of the index $b$. The observations that place the tightest limits are
those with the highest energies and largest path lengths. For example,
10 -- 20 TeV photons have been detected from Mkn 421 and Mkn 501 [11]
at redshifts of $z=$ 0.031 and 0.033 ($\dist$ $\approx$ 140 Mpc).
Many other sources have been observed with photon energies $\egam$ =
0.3 -- 10 TeV and distances $\dist$ = 100 -- 500 Mpc [12]. In Figure
2, this set of observations is depicted as the dark band in the
$b-\mpl$ plane. The region below the band is ruled out, whereas the
region above the band remains viable.


The bounds discussed in this paper require four conditions: (1) The
vacuum is described by the paradigm of space-time foam, where virtual
black holes flicker in and out of existence with a mean density of one
virtual black hole per Planck volume.  (2) The virtual black holes
driving this effect do not preserve the identity of the photons they
absorb. (3) The cross section for absorption followed by two particle
emission has the assumed form (eq. [\ref{eq:crossb}]) in the long
wavelength limit, which requires that (4) Lorentz invariance is
violated (or modified [6,7]) at the Planck scale. 

Existing astronomical observations already constrain theories of
quantum gravity. The results shown in Figures 1 and 2 indicate that
quantum gravity is constrained by at least one of the following
conditions: (A) The Planck mass must be relatively large ($\mpl >
10^{15}$ GeV), or (B) The interaction cross sections must be highly
suppressed over their expected values (either $b \gg 5$ or $\beta_2
\ll 1$), or (C) Quantum gravity does not violate (or modify) standard
Lorentz invariance.  Future astronomical observations will probe more
of the parameter space for which quantum foam can affect photon
propagation and will thereby provide even tighter limits.
Observations of high energy sources ($\egam > 20$ TeV) out to greater
distances ($d > 200$ Mpc) will provide the first new constraints. This
type of observation will be limited when photon energies approach the
threshold at 300 TeV due to interactions with CMB photons. To make
further progress, extremely high energies ($\egam > 10^8$ GeV) are
needed.



{\bf Acknowledgements:} We especially thank Malcolm Perry for
preliminary discussions that led to this paper. We also thank
C. Akerlof, P. Bucksbaum, G. Kane, F. Larsen, J. Liu, Greg Tarl{\'e},
and J. Wells for additional discussion. This work was supported by the
University of Michigan, in part through the Michigan Center for
Theoretical Physics.
 

\newpage 
\baselineskip=16pt 
\parskip=1pt 

{\bf References}  

\medskip\par\pp{[1]} 
J. Wheeler, {\sl Ann. Phys.} {\bf 2} (1957) 604; 
S. Hawking, {\sl Comm. Math. Phys.} {\bf 87} (1982) 395. 

\medskip\par\pp{[2]} 
S. W. Hawking, {\sl Nucl. Phys.} {\bf B144} (1978)  349;
G. W. Gibbons and S. W. Hawking, {\sl Phys. Rev.} D {\bf 15} (1977) 2752; 
S. W. Hawking, D. N. Page, and C. N. Pope, {\sl Phys. Lett.} {\bf B86} 
(1979) 175. 

\medskip\par\pp{[3]} 
R. C. Myers and M. J. Perry, {\sl Ann. Phys.} {\bf 172} (1986) 304; 
F. R. Tangherlini, {\sl Nuovo Cimento} {\bf 27} (1963) 636; 
M. J. Perry, in {\sl Unification of Elementary Forces and Gauge Theories}, 
ed. D. B. Cline and F. E. Mills (Harwood, London, 1977) p. 485. 

\medskip\par\pp{[4]} 
S. W. Hawking, {\sl Comm. Math. Phys.} {\bf 43} (1975) 199;  
{\sl Nature} {\bf 248} (1974) 30.  

\medskip\par\pp{[5]}  
D. N. Page, {\sl Phys. Rev.} D {\bf 13} (1976) 198; 
A. A. Starobinsky and S. M. Churilov, {\sl Zh. Eksp. Teor. Fiz.} 
{\bf 65} (1973) 3 [{\sl Sov. Phys. JETP} {\bf 38} (1974) 1]. 

\medskip\par\pp{[6]} 
T. Jacobson, S. Liberati, and D. Mattingly, {\sl Phys. Rev.} D, 
submitted, hep-ph/0209264 (2002); 
T. Jacobson, S. Liberati, and D. Mattingly, {\sl Phys. Rev.} D 
{\bf 66} (2002) 081302.   

\medskip\par\pp{[7]}
G. Amelino-Camelia et al., {\sl Int. J. Mod. Phys.} A {\bf 12} (1997) 607; 
D. Colladay and V. A. Kosteleck{\'y}, {\sl Phys. Rev.} D {\bf 58}
(1998) 116002; G. Amelino-Camelia et al., {\sl Nature} {\bf 393}
(1998) 763; G. Amelino-Camelia, gr-qc/0201012 (2002). 

\medskip\par\pp{[8]} 
D. N. Spergel et al., submitted to {\sl Astrophys. J.} (2003) 
astro-ph/0302209. 

\medskip\par\pp{[9]} 
I. Antoniadis, {\sl Phys. Lett.} {\bf B246} (1990) 377; 
J. Lykken, {\sl Phys. Rev.} D {\bf 54}, hep-th/9603133 (1996) 3693; 
K. R. Dienes, E. Dudas, and T. Gherghetta, {\sl Phys. Lett.} {\bf B436} 
(1998) 55; P. C. Argyres, S. Dimopoulos, and J. March-Russell, 
{\sl Phys. Lett.} {\bf B441} (1998) 96; J. Ellis, A. E. Faraggi, 
and D. V. Nanopoulos, {\sl Phys. Lett.} {\bf B419} (1998) 123; 
N. Arkani-Hamed, S. Dimopoulos, and G. Dvali, {\sl Phys. Lett.}  
{\bf B429} (1998) 263; {\sl Phys. Rev.} D {\bf 59} (1999) 0806004; 
I. Antoniadis, N. Arkani-Hamed, S. Dimopoulos, and G. Dvali, {\sl
Phys. Lett.} {\bf B436} (1999) 257.

\medskip\par\pp{[10]} 
F. C. Adams, G. L. Kane, M. Mbonye, and M. J Perry, 
{\sl Int. J. Mod. Phys.} {\bf A 16} (2001) 2399. 

\medskip\par\pp{[11]} 
F. Aharonian et al., {\sl Astron. Astrophys.} {\bf 349} (1999) 11.  

\medskip\par\pp{[12]} 
M. Catanese and T. C. Weekes, {\sl PASP} {\bf 111} (1999) 1193.  

\medskip\par\pp{[13]} 
S. Sarkar, {\sl Mod. Phys. Lett.} {\bf A17} (2002) 1025; 
F. Stecker, 2nd VERITAS Symposium on TeV Astrophysics, 
astro-ph/0304527; J. R. Ellis et al., {\sl A \& A} {\bf 402} (2003) 409. 

\medskip\par\pp{[14]} 
P.J.E. Peebles, {\sl Principles of Physical Cosmology} 
(Princeton Univ. Press, Princeton, 1993) p. 182. 


\newpage 
\begin{figure}
\figurenum{1}
{\centerline{\epsscale{0.90} \plotone{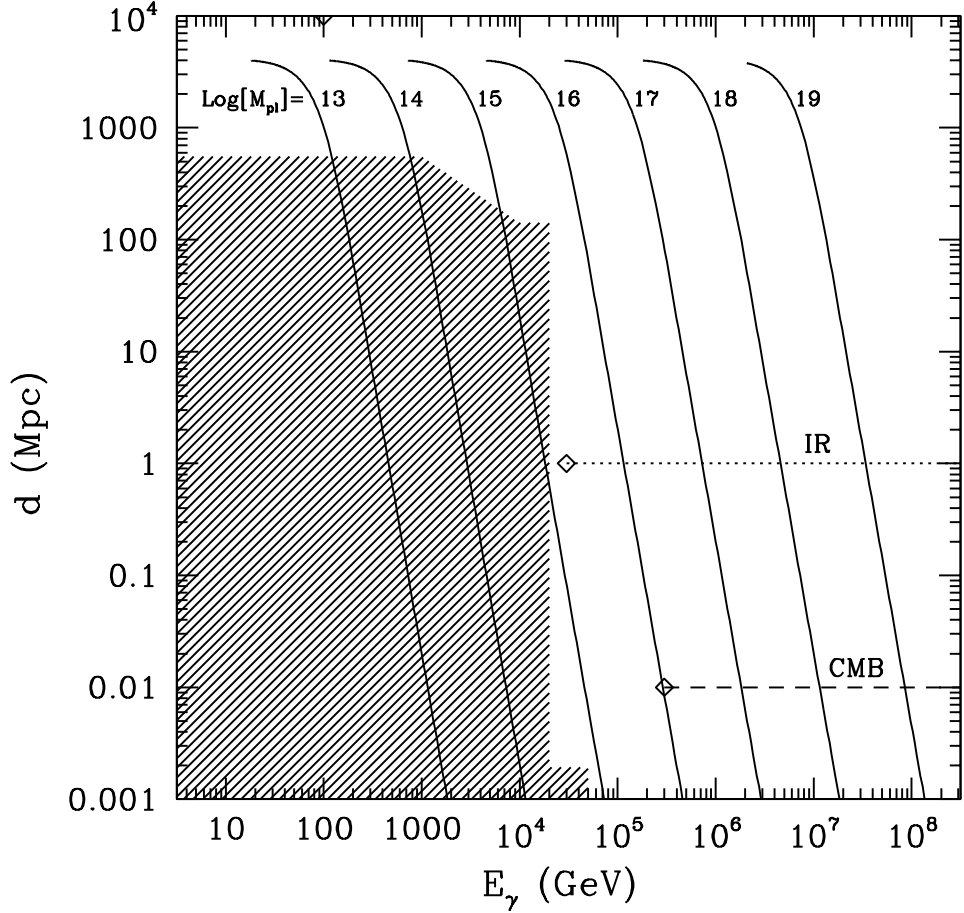} }} 
\figcaption{The predicted quantum foam cutoff as a function of photon
energy for varying values of the Planck mass. The maximum distance
$\dist$ for which astrophysical photons can propagate is shown as a
function of observed photon energy $\egam$ (i.e., present-day energy).  
The cross section is assumed to have the form given by equation 
(\ref{eq:crossb}) with $b$ = 5. The Planck mass varies from $10^{13}$ 
to $10^{19}$ GeV, as indicated near the top of each curve. The shaded 
region shows the portion of the plane that has been probed by 
astronomical observations of high energy photons (see text).  The
dashed curve labeled CMB shows the maximum path length due to
scattering of high energy photons by the cosmic microwave background;
this cutoff at 10 kpc operates for photon energies $\egam > 3 \times
10^5$ GeV.  The dotted curve labeled IR shows the maximum path length
due to the infrared background; this cutoff at 1 Mpc operates for
photon energies $\egam > 3 \times 10^4$ GeV. }  
\end{figure}

\newpage 
\begin{figure}
\figurenum{2}
{\centerline{\epsscale{0.90} \plotone{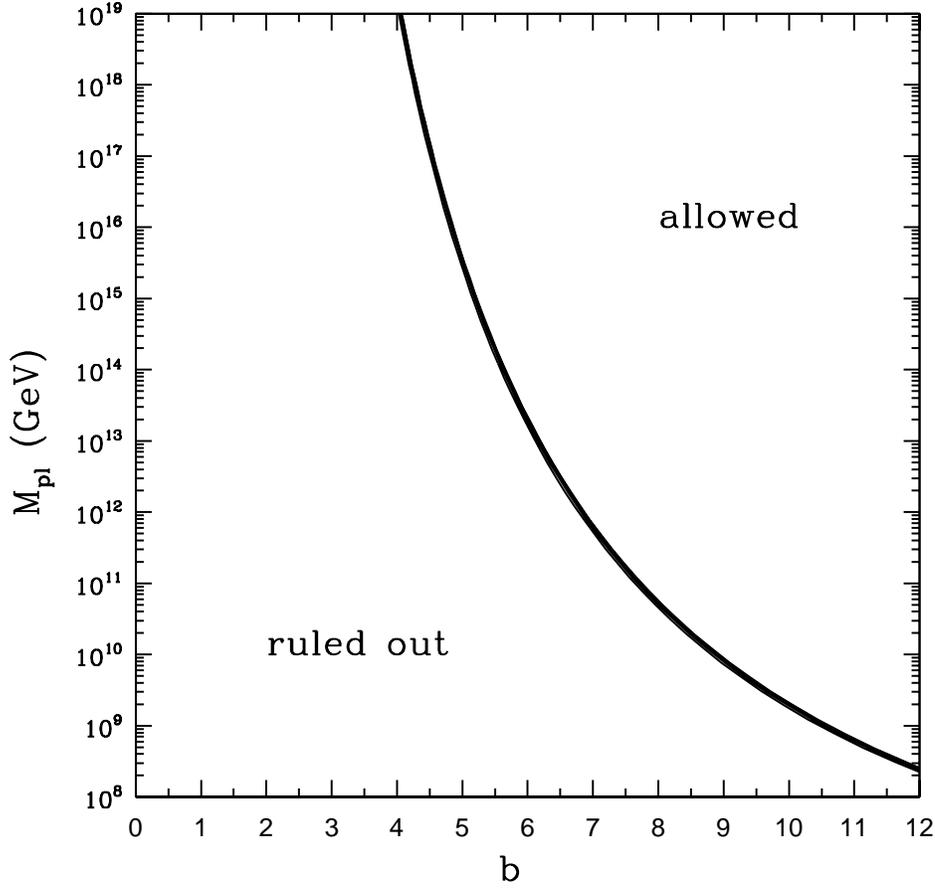} }} 
\figcaption{Constraints on the $b-\mpl$ plane from the quantum foam
cutoff. The horizontal axis corresponds to the index $b$ that appears
in the interaction cross section; the vertical axis shows the Planck
mass $\mpl$ (which can be lower than the standard value).  Existing
observations of high energy photons from extragalactic sources
constrain the possible interactions between photons and virtual black
holes. The dark band depicts the region of the plane probed by
observations with photon energies $\egam$ = 20 TeV and source 
distances $d$ = 100 -- 500 Mpc.  The allowed region of the plane is
above the curves on the upper right; the region to the lower left of
the curves is ruled out. }

\end{figure}

\end{document}